\makeatletter
\def\@maketitle{%
\defaultfont\normalsize
\let\@makefnmark\relax \let\@thefnmark\relax \ifx\@empty\@subjclass\else
\@footnotetext{1991 {\it Mathematics Subject
Classification}.\enspace
\@subjclass.}\fi
\ifx\@empty\@keywords\else
\@footnotetext{{\it Key words and phrases.}\enspace \@keywords.}\fi
\ifx\@empty\@thanks\else
\@footnotetext{\@thanks}\fi
\topskip66\p@ 
\vtop{\centering{\baselineskip14\p@\bf
\expandafter{\@title}\@@par}%
\global\dimen@i\prevdepth}%
\prevdepth\dimen@i
\ifx\@empty\@authors
\else
\baselineskip32\p@
\vtop{\@andify{ AND }\@authors
\centering{{\@authors}\@@par}%
\global\dimen@i\prevdepth}\relax
\prevdepth\dimen@i
\fi
\ifx\@empty\@dedicatory
\else
\baselineskip18\p@
\vtop{\centering{\small\it\@dedicatory\@@par}%
\global\dimen@i\prevdepth}\prevdepth\dimen@i \fi
\ifx\@empty\@date\else
\baselineskip24\p@
\vtop{\centering\@date\@@par
\global\dimen@i\prevdepth}\prevdepth\dimen@i \fi

\normalsize
\dimen@32\p@ \advance\dimen@-\baselineskip \vskip\dimen@\@plus14\p@
} 
\makeatother

\documentstyle[amscd,amssymb,verbatim,12pt]{amsart}

\theoremstyle{plain}
\newtheorem{Thm}{Theorem}[section]
\newtheorem{Cor}[Thm]{Corollary}
\newtheorem{Lem}[Thm]{Lemma}
\newtheorem{Prop}[Thm]{Proposition}
\newtheorem{Def}[Thm]{Definition}

\theoremstyle{remark}
\newtheorem{Ex}[Thm]{Example}

\newcommand{\ra}{\rightarrow}

\newcommand{\Der}{\operatorname{Der}}

\newcommand{\pvf}{\operatorname{pvf}}
\newcommand{\vf}{\operatorname{vf}}
\def\pv#1{#1-$\pvf{}$}
\def\morph#1{\overset{#1}{\ra}}

\def\dsp#1{$\displaystyle{#1}$}
\def\Cal#1{{\cal #1}}
\def\BBB#1{{\Bbb #1}}

\def\Frak #1{{\frak{#1}}}

\theoremstyle{remark}

\numberwithin{equation}{section}

\newcommand{\Oplus}{\operatornamewithlimits{\oplus}}

\newcommand{\C}{\BBB C}

\title[ZCC]{Parametrized vector fields and the zero-curvature condition}

  \author{ Mitchell Rothstein}

\begin{document}
\begin{abstract}We apply the notion of a parametrized vector field on a manifold $M$,  where the parameters are also in $M$,  to the study of the zero-curvature condition that arises in the context
of integrable systems.
\end{abstract}

\maketitle
\section{Motivation}
Consider the following result,  which is perhaps well-known to experts.
 Let $R$ be a ring of characteristic zero,  equipped with a derivation $\partial$,  and let
\begin{equation}
\label{eq:psdo}
L=\partial +\sum_{i=-\infty}^{-1} f_i \partial^i
\end{equation}
be a first-order monic formal
pseudodifferential operator with coefficients in   $R$.    Let $\Psi_L$ denote the ring of pseudodifferential operators commuting with $L$.   Given  positive integers $n,m$,  there is a Lie algebra homomorphism
\begin{align}
\label{eq:map.1}
\Psi_L&\to \Psi_L[[s,t]]\\
K&\mapsto K(s,t)
\end{align}
given by solving the initial value problem
\begin{align}
  \label{eq:kpflows.1}
\frac{\partial K}{\partial s}&=[L^n_+,K]\label{flow1.1}\\
\frac{\partial K}{\partial t}&=[L^m_+,K]\label{flow2.1}\\
K(0,0)&=K\ .
\end{align}
In particular,  
flows \eqref{flow1.1} and \eqref{flow2.1}
commute.   This appears paradoxical at first,   since $L^n_+$ and $L^m_+$ do not,  in general,  commute.    The resolution of the paradox is that  $L^n_+$ and $L^m_+$ are not fixed operators,  but are themselves subject to the flows \eqref{flow1.1} and \eqref{flow2.1}.

Better known is the special case $K=L$,   $R=\BBB C[[x]]$.   Indeed,  one then has the family of KP flows,  whose commutativity may be understood from  various points of view.  (
\cite{A,SW}).  

Here is a somewhat sharper statement,  which may not be so well-known:    Let $L_1$ and $L_2$ be commuting pseudodifferential operators.   Let $\Psi_{L_1,L_2}$ 
denote the ring of pseudodifferential operators commuting with $L_1$ and $L_2$.   Given  positive integers $n,m$,  there is a Lie algebra homomorphism
\begin{align}
\label{eq:map}
\Psi_{L_1,L_2}&\to \Psi_{L_1,L_2}[[s,t]]\\
K&\mapsto K(s,t)
\end{align}
given by solving the initial value problem
\begin{align}
  \label{eq:kpflows}
\frac{\partial K}{\partial s}&=[{L_1}_+,K]\label{flow1}\\
\frac{\partial K}{\partial t}&=[{L_2}_+,K]\label{flow2}\\
K(0,0)&=K\ .
\end{align}

The assumption  that $K$ commutes with $L_1$ and $L_2$  is not fundamental.  Its role is to guarantee that the order of $K$ remains bounded.
Issues of this sort arise in the infinite dimensional setting.    Indeed,   the commutativity of the flows  \eqref{flow1} and \eqref{flow2}  holds with $\Psi_{L_1,L_2}$ replaced by an arbitrary finite dimensional Lie algebra. (Corollary \ref{cor:consistent}.)

The purpose of this note is to give an understanding of the commutativity of the flows \eqref{flow1} and \eqref{flow2}  in terms of parametrized vector fields on a manifold,  where the  parameters are themselves elements of the manifold.   Given such a vector field,    and given a choice of the parameters,   one obtains a flow,   with the understanding that  the parameters themselves also flow.    The next section describes this setup.    We specialize to Lie algebras in the subsequent section.
To avoid irrelevant complications,  we will work with finite dimensional manifolds.   Throughout the paper,    $M$ denotes  such a manifold.

\section{\pv n's}

\begin{Def}
  \label{parametrized vector field}   Given a positive integer $n$,  let $\pvf_n(M)$ denote the space of sections
\begin{equation}
  \label{eq:parameterized vector field}
 \xi:M^{n+1}\to p_{n+1}^*(TM)\ , 
\end{equation}
where  ${p_i}$ is projection onto the $i^{th}$ factor and $TM$
is the  tangent  bundle.     

We refer to the elements of $\pvf_n(M)$ as \pv n's. \end{Def}

One  may think of an \pv n as an object which,   for every choice of $a_1,...,a_{n}\in  M$  {determines a flow on $M$} in the following way.   Given $b\in M$,  one has a tangent vector $\xi(a_1,...,a_{n},b)$ to $M$ at $b$.    Move infinitesimally in the direction of that vector,  to a nearby point $b'$.    Having done this for all $b$,  one has done it in particular for $a_1,...,a_{n}$,   so at the next iteration,  move in the direction $\xi(a_1',...,a_{n}',b')$.    

To put this more precisely, let $\vf(M^{n})$ denote the space of vector fields  on $M^{n}$.  For $i=1,...,n$,  set
\begin{align}
\tau_i:M^{n}&\to M^{n}\\
\tau_i(a)&=(a_1,...,a_{n-1},a_i)\ ,
\end{align}
where $a=(a_1,...,a_{n})$.
Taking account of the natural isomorphism  \begin{equation}\label{eq:directSum}
T(M^n)=\Oplus_i {p_i}^*(TM)\  ,
\end{equation}
define
\begin{align}
  \label{eq:right inverse}
\pvf_{n-1}(M)&\overset\beta\to\vf(M^{n})\\
\xi&\mapsto\tilde \xi\ ,
\end{align}
by
\begin{equation}
  \label{eq:right inverse 2}
  \tilde\xi(a_1,...,a_{n})=(\xi\circ\tau_1(a),...,\xi\circ\tau_{n}(a))\  .
\end{equation}

Note that  $\beta$ has a left inverse,
\begin{equation}
  \label{eq:leftt inverse}
\vf(M^{n})\overset{\pi_n}\to\pvf_{n-1}(M)
\end{equation}
given by projection onto  the $n^{th}$ summand in equation \eqref{eq:directSum}.

\begin{Prop}\label{prop;commmuteWithTau}
 $\beta$ maps $\pvf_{n-1}(M)$ isomorphically onto the Lie subalgebra
$$\{\ \eta\in\vf(M^{n})\ |\ \forall i\ \tau_i^*\circ\eta=\eta\circ\tau_i^*\ \}\ .
$$
Thus  $\pvf_{n-1}(M)$  forms a Lie algebra,  with bracket
\begin{equation}
\label{eq:(n-1)-bracket}
[\xi,\psi]=\pi_n([\tilde\xi,\tilde\psi]) \ .
\end{equation}
\end{Prop}

\begin{pf}  Note that if $k\ne n$,
\begin{align}
 \tau_i^* \circ p_k^*  &= p_k^* \label{one} \\
   \tau_k \tau_i&=  \tau_k\label{one} 
\end{align}
while
\begin{align}
 \tau_i^* \circ p_n^*  &= p_i^* \label{three} \\
   \tau_n \tau_i&=  \tau_i\label{four} \ .
\end{align}
Let $\xi\in\pvf_{n-1}(M)$.   Let $f\in C^{\infty}(M)$.     Then for all $k$,
\begin{equation}
\label{ eq;explicitFormula}
\tilde\xi\circ p_k^*(f)|_a=\xi|_{\tau_k(a)}(f)\ .
\end{equation}
If $k\ne n$,
\begin{align}
\tilde\xi\circ\tau_i^*\circ p_k^*(f)|_a&=\xi|_{\tau_k(a)}(f)=\notag\\
\xi|_{\tau_k\tau_i(a)}(f)&=\tau_i^*\circ\tilde\xi\circ p_k^*(f)|_a\notag
\end{align}
while
\begin{align}
\tilde\xi\circ\tau_i^*\circ p_n^*(f)|_a&=\xi|_{\tau_i(a)}(f)=\notag\\
\xi|_{\tau_n\tau_i(a)}(f)&=\tau_i^*\circ\tilde\xi\circ p_n^*(f)|_a\ .\notag
\end{align}
Thus,   $ \tau_i^*\circ\tilde\xi=\tilde\xi\circ\tau_i^*$ for all $i$.

Conversely,  let $\eta$ be a vector field such that $ \tau_i^*\circ\eta=\eta\circ\tau_i^*$ holds for all $i$.  Let $\eta_n$ denote the $n^{th}$ component of $\eta$. Then
\begin{align}
\eta\circ p_i^*(f)|_a&=\eta\circ\tau_i^*\circ p_n^*(f)|_a=\notag\\
\tau_i^*\circ\eta\circ p_n^*(f)|_a&=\eta_n|_{\tau_i(a)}(f)\notag\\
&=\widetilde{\pi_n(\eta)}\circ p_i^*(f)|_a\ .\notag
\end{align}
Thus $\eta=\widetilde{\pi_n(\eta)}$.
\end{pf}

\begin{Ex}   If $M=\Bbb R$,  then a \pv 1 is simply
a function $f(x,y)$.   Then 
\begin{equation}
\tilde f= f(x,x)\frac{\partial}{\partial x}+f(x,y)\frac{\partial}{\partial
y}
\end{equation} Then the  bracket of  \pv 1's is given by
\begin{equation}
[f,g]=f(x,x)\frac{\partial g}{\partial x}+f(x,y)\frac{\partial g}{\partial
y}-
g(x,x)\frac{\partial f}{\partial x}-g(x,y)\frac{\partial f}{\partial
y}\end{equation}
\end{Ex} 

Now let $\xi$ be a \pv 1.    
Introduce the following \pv 2's:
\begin{align}
\xi^{1}(a,b,c)&=\xi(a,c)\\
\xi^{2}(a,b,c)&=\xi(b,c)
\end{align}

\begin{Def}[Zero-curvature condition]
  \label{zcc}
Fix a  \pv 1, $\xi$,  on $M$.   A pair of points  
 $a,b\in M$ satisfies the zero-curvature condition  (for $\xi$),
zcc,   if, for all $y\in M$,
\begin{equation}
  \label{eq:zcc}
  \Phi^{1}_s\Phi^{2}_t(a,b,y)=\Phi^{2}_t\Phi^{1}_s(a,b,y)\ ,
\end{equation}
where $\Phi^{i}_{\cdot}$ is the flow of  $\widetilde{\xi^{i}}$.
\end{Def}
In other words,  $(a,b)$ satisfies zcc if the flow determined by 
$b$ commutes with the flow determined by $a$.

\subsection{ Infinitesimal criterion}

Let $\Frak F_n$ denote the free Lie algebra on $n$ letters 
$\alpha_1,...,\alpha_n$.  Let $\Frak I_n\subset \Frak F_n$ denote the commutator ideal.   Given $w\in \Frak I_n$, and given elements $x_1,...,x_n$ in a Lie algebra $\Frak g$,  let
$w(x_1,...,x_n)\in\Frak g$ denote the element obtained by evaluating $\alpha_i$ at $x_i$.

%

For a vector field $X$ on  $M$,  denote by $\Phi^X$ the one-parameter group of diffeomorphisms generated by $X$.

\begin{Lem}
Let $X$ and $Y$ be smooth vector fields on  $M$, and let $p\in M$.   
If there is a neighborhood $(0,0)
\in U
\subset \BBB R^2$ such that for all $(s,t)\in U$,
\begin{equation}
  \label{eq:commuting flows}
  \Phi_t^Y\Phi_s^X(p)=\Phi_s^X\Phi_t^Y(p)\ ,
\end{equation}
then
\begin{equation}
  \label{eq:vanishing}
 \forall w\in\Frak I_2\  ,\   w(X,Y)(p)=0\ .
\end{equation}
Conversely, if $X$ and $Y$ are analytic vector fields on an analytic
manifold $M$, then \eqref{eq:vanishing} implies \eqref{eq:commuting
flows}.
\end{Lem}

\begin{pf}
Assume \eqref{eq:commuting flows}.    
Let
\begin{align}
  \label{eq:surface}
U&\morph\alpha M\\
\alpha(s,t)&=\Phi_s^X\Phi_t^Y(p)\ .
\end{align}
From the two sides of  \eqref{eq:commuting flows} respectively, one has 
\dsp{\frac{\partial \alpha}{\partial t}=Y|_{\alpha}} and  
\dsp{\frac{\partial \alpha}{\partial s}=X|_{\alpha}}.    It follows that
if $X$ and $Y$ are linearly  independent at $p$, 
one may choose coordinates $(x_1,...,x_n)$ centered at $p$,   such that 
\begin{align}
  \label{eq:straight}
X&=\frac{\partial}{\partial x_1}+\tilde X\\
Y&=\frac{\partial}{\partial x_2}+\tilde Y \ ,
\end{align}
where $\tilde X$ and $\tilde Y$ vanish along $x_2=x_3=...=x_n=0$.  Then 
\eqref{eq:vanishing} holds.  It is also clear that  \eqref{eq:vanishing}
holds if both $X$ and $Y$
 vanish at $p$.

It remains to consider the case that  $X$ is nonvanishing in a neighborhood of
 $p$, and $Y(p)$ is a multiple of $X(p)$. 
 Note that equation \eqref{eq:commuting flows} 
holds with $p$ replaced by
$\alpha(s,t)$.    Therefore, 
if there is a sequence $s_k\ra 0$, such that $X(\alpha(s_k,0))$ 
and $Y(\alpha(s_k,0))$
are linearly independent, \eqref{eq:vanishing} will hold at $p$, by continuity.
The remaining possibility is that $Y$ is a multiple of $X$ at $\alpha(s,0)$ 
for all $s$ in a neighborhood of $0$.   Then we may assume
\begin{align}
  \label{eq:straighter}
X&=\frac{\partial}{\partial x_1}\\
Y&=f(x_1,...,x_n)\frac{\partial}{\partial x_1}+\tilde Y \ ,
\end{align}
where  $\tilde Y$ vanish along $x_2=x_3=...=x_n=0$.   This reduces to the case
that $M=\BBB R^1$, \dsp{X=\frac{d}{ dx}},  
\dsp{Y=f(x)\frac{d}{ dx}} and $p=0$.   Let $\gamma(t)$ be the integral curve 
of $Y$ with $\gamma(0)=0$.   Then 
\begin{equation}
  \label{eq:linear1}
  \Phi_{t_1}^Y(\gamma(t_2))=\Phi_{t_1}^Y\Phi_{t_2}^Y(0)=\gamma(t_1+t_2)\ .
\end{equation}
Now
\begin{align}
  \label{eq:linear2}
\Phi_{t_1}^Y\Phi_{\gamma(t_2)}^X(0)&=\Phi_{t_1}^Y(\gamma(t_2))=\gamma(t_1+t_2)
\\
=\Phi_{\gamma(t_2)}\Phi_{t_1}^Y(0)&=\Phi_{\gamma(t_2)}^X(\gamma(t_1))=
\gamma(t_1)+\gamma(t_2)\ .
\end{align}
Then there exists a constant,  $c$,  such that \dsp{Y=c\frac{d}{ dx}}, so  
\eqref{eq:vanishing} holds.

Conversely,  in the analytic setting,  the Baker-Campbell-Hausdorff formula 
furnishes a set of elements $w_{i,j}\in\Frak I_2$ with the following property:
Given any analytic function $f$ in a neighborhood of $p$, 
\begin{equation}
  \label{eq:bch}
  f(\Phi_{-s}^X\Phi_{-t}^Y\Phi_{s}^X\Phi_{t}^Y(p))=\sum_{k=0}^{\infty} \frac 1{k!}(\sum_{i,j>0}s^it^jw_{i,j}(X,Y))^k(f)|_p\ .
\end{equation}
Thus \eqref{eq:vanishing} implies \eqref{eq:commuting flows}
\end{pf}

Thus  one has the following infinitesimal criterion.

\begin{Cor}\label{cor:infinitesimal}  Fix an analytic \pv 1 $\xi$ on
 an
analytic manifold $M$.    Then zcc holds for $(a,b)$ if and only if , for all $w\in\Frak I_2$, 
$w(\xi^1,\xi^2)(a,b,z)=0$ for all $z\in M$.

\end{Cor}

\section{The case of a Lie algebra}

Let $\Frak g$ be a finite dimensional Lie algebra.
An  \pv n on $\Frak g$ is simply a smooth function $\xi:\Frak g^{n+1}\to \Frak g$.
In particular,  for all smooth $f:\Frak g^n\to \Frak g$, consider the 
\pv n
\begin{equation}\label{eq:Laxpair}
\xi_f(x_1,...,x_{n},y)=[f(x_1,...,x_{n}),y]\  .
\end{equation}
We will say that such an \pv n is of {\em Lax type}.
\begin{Prop}\label{prop:Laxbracket}
The \pv n's of Lax type 
form a lie subalgebra of the 
\pv ns.   More precisely, set $x=(x_1,...,
x_n)$ and set $[f(x),x]=([f(x),x_1],...,[f(x),x_{n}])$.  Then
\begin{equation}\label{eq:Laxbracket}
[\xi_f,\xi_g]=\xi_{[f,g]'}\ ,\end{equation}
where
\begin{equation}
\label{eq:functionbracket}
[f,g]'(x)=dg_x([f(x),x])-df_x([g(x),x]+[g(x),f(x)] \ .\end{equation}
\end{Prop}

\begin{pf}
\begin{equation}
[\xi_f,\xi_g](x,y)=\frac d{dt}|_0([g(x+t[f(x),x]),y+t[f(x),y]]- (f\leftrightarrow g)
)\ .\end{equation}
Then use the Jacobi identity.
\end{pf}

\begin{Prop}\label{p:Lie algebra}
 Equation \eqref{eq:functionbracket} endows 
$C^{\infty}(\Frak g^{n};\Frak g)$ with the structure of a Lie algebra.
\end{Prop}

\begin{pf}
It is not difficult to prove the proposition by direct calculation.
A more conceptual proof is as follows.
 Given a manifold  $M$  equipped with an infinitesimal 
$\Frak g$-action,
$\nabla:\Frak g\to \Der(C^{\infty}(M))$,  the space 
$C^{\infty}(M)\otimes\Frak g$ has a natural  Lie algebra  structure,  $[\cdot,\cdot]'$, given by
\begin{equation}
\label{ bracket}
[a\otimes X,b\otimes Y]'=a\nabla_X(b)\otimes Y-b\nabla_Y(a)\otimes X+ab\otimes[X,Y]\ . 
\end{equation}
There is a natural infinitesimal action of  $\Frak g$ on $\C^{\infty}(\Frak g^ n)$,  induced by the 
coadjoint action.   The resulting bracket is precisely \eqref{eq:functionbracket},  up to sign.
\end{pf}


\section{Main Result}  

 Note that  $C^{\infty}(\Frak g^n,\Frak g)$  now has two Lie algebra structures,   the pointwise bracket and the bracket  given by  \eqref{eq:functionbracket} .     Denote these two Lie algebras by $C^{\infty}(\Frak g^n,\Frak g)^P$ and $C^{\infty}(\Frak g^n,\Frak g)'$ respectively.  
  
 Though the  next theorem shares the hypothesis of the AKS theorem, \cite{A},   it  seems not to be a corollary.

\begin{Thm}\label{AKSlike}   Let $\Frak g=\Frak g_+\oplus \Frak g_-$ be a
vector space direct sum decomposition of $\Frak g$,   such that  $\Frak
g_{\pm}$ are Lie subalgebras.  Let
$\Frak G^P\subset C^{\infty}(\Frak g^n,\Frak g)^P$ denote the Lie 
subalgebra generated by the projections
${p_1},...,{p_n}$.    Let $\Frak G'\subset C^{\infty}(\Frak 
g^n,\Frak g)'$ denote the Lie
subalgebra generated by ${p_1}_+,...,{p_n}_+$,   where $(\cdot)_+ $ 
denotes projection onto
$\Frak g_+$.    Then the map of vector spaces
\begin{align}
C^{\infty}(\Frak g^n,\Frak g)&\to  C^{\infty}(\Frak g^n,\Frak g)\notag\\
f&\mapsto f_+
\end{align}
restricts to a Lie algebra homomorphism
\begin{equation}
\Frak G^P\to \Frak G'\ .\end{equation}
\end{Thm}
\begin{pf} As before,  let $\Frak F_n$ denote the free Lie algebra on $n$ letters 
$\alpha_1,...,\alpha_n$.   
We must prove that for all $\omega\in \Frak F_n$,
\begin{equation}\label{wordfmla}
\omega(p_1,...,p_n)_+=\omega({p_1}_+,...,{p_n}_+)\  .
\end{equation}
Given a word $w=b_1b_2...b_k$ over the alphabet $\alpha_1,...,\alpha_n$,  let
\begin{equation}\label{standardmonomial}
\tilde w=[b_1,[b_2,[...[b_{k-1},b_k]]...]\ .\end{equation}
The words $\tilde w$ span $\Frak F_n$, so it suffices to
establish \eqref{wordfmla} when
$\omega=\tilde w$.    Fix an index $i$, and consider the set of functions $f\in C^{\infty}(\Frak g^n,\Frak
g)$ satisfying the following equation:
\begin{equation}\label{covariantderivative}
df([{p_i}_+,p])=[{p_i}_+,f]\ ,\end{equation}
where $p=({p_1},...,{p_n})$.   It is clear that this set of  functions
forms a Lie subalgebra of
$C^{\infty}(\Frak g^n,\Frak g)^P
$.   Furthermore,   for all  $j$,   \eqref{covariantderivative}
holds when $f=p_j$.
Thus,  \eqref{covariantderivative} holds for all $f\in\Frak G^P$.
  Given
\eqref{covariantderivative},  one has
\begin{align}
[{p_i}_+,f_+]'&=df_+([{p_i}_+,p])-{p_i}_+([f_+,p])+[f_+,{p_i}_+]\notag\\
&=[{p_i}_+,f]_+-[f_+,{p_i}]_+-[{p_i}_+,f_+]\notag\\
&=[{p_i}_+,f_-]_+-[f_+,{p_i}]_+\notag\\
&=[{p_i},f_-]_+-[f_+,{p_i}]_+\notag\\
&=[p,f]_+\ .
\end{align}
Then \eqref{wordfmla} holds for all words,  by induction on the length.

\end{pf}

As a corollary,  one finds that when $a$ and $b$ are  commuting elements
of $\Frak g$,   the flows
\begin{align}
  \label{eq:commutingflows}
\frac{\partial c}{\partial s}&=[a_+,c]\\
\frac{\partial c}{\partial t}&=[b_+,c]\\
\end{align}
commute,  irrespective of any functional dependence among $a$, $b$ and $c$.
This is made precise in the following corollary.

\begin{Cor} \label{cor:consistent} With $\Frak g$ as in theorem \ref{AKSlike}, consider the
\pv 1
\begin{equation}
\xi(x,y)=[x_+,y]\ .\end{equation}
Then a pair of elements $(a,b)\in\Frak
g\times\Frak g$ satisfies zcc is and only if, for all $w\in\Cal I_2$,
$w(a,b)_+$ belongs to the center of $\Frak g$.

In particular,  for $a$ and $b$  to satisfy zcc it is sufficient that $a$ and $b$ commute. 
\end{Cor}

\begin{pf}
By corollary \ref{cor:infinitesimal} ,  the necessary and sufficient condition is that for all $w\in\Cal I_2$ and all $z\in \Frak g$,  $w(\xi^1,\xi^2)(a,b,z)=0$.    Now $\xi^1$ and $\xi^2$ are the Lax type \pv 2's $\xi_{{p_1}_+}$ and  $\xi_{{p_2}_+}$
respectively.   Then,  by proposition \ref{prop:Laxbracket} and theorem \ref{AKSlike}
\begin{equation}
w(\xi^1,\xi^2)=\xi_{w({p_1}_+,{p_2}_+)}=\xi_{w({p_1},{p_2})_+}\  .
\end{equation}
 
\end{pf}

This corollary recovers the commutativity of the flows \eqref{flow1} and \eqref{flow2}.

We conclude with a finite dimensional example.

\section{Example}    Let $\Frak g=\Frak {sl}_3(\Bbb R)$.   Let $\Frak g_+$ be the
subalgebra of skew-symmetric matrices and let $\Frak g_-$
be the
subalgebra of upper triangular matrices.   Take
\begin{equation}
a
=
\left [\begin {array}{ccc}
0&0&0\\\noalign{\medskip}0&0&0
\\\noalign{\medskip}1&0&0\end {array}\right ]\ ;\ 
b=
\left [\begin {array}{ccc} 0&0&0\\\noalign{\medskip}0&0&0
\\\noalign{\medskip}0&1&0\end {array}\right ]
\end{equation}
First we solve the initial value problem
\begin{align}
\frac{d a(s)}{d s}&=[a(s)_+,a(s)]\\
a(0)&=a\ .
\end{align}
By developing the solution in a power series about $s=0$, one finds 
the following solution.
\begin{equation}
a(s)=\left [\begin {array}{ccc} -{\frac {s}{1+{s}^{2}}}&0&-{\frac
{{s}^{2}}
{1+{s}^{2}}}\\\noalign{\medskip}0&0&0\\\noalign{\medskip}{\frac {{1}
}{1+{s}^{2}}}&0&{\frac {s}{1+{s}^{2}}}\end {array}\right ]
\end{equation}
Now  the problem
\begin{align}
\frac{d b(s)}{d s}&=[a(s)_+,b(s)]\\
b(0)&=b\ .
\end{align}
can be solved by the dressing method.   That is,  we  let $\sigma(s)\in SL(3,\BBB C[[s]])$ be the solution of the initial value problem
\begin{align}
\label{eq;ivp}
    \frac{d\sigma(s)}{ds}\sigma(s)^{-1}&=a(s)_+   \\
    \sigma(0)  &=I\ .
\end{align}
Then 
\begin{equation}
b(s)=ad_{\sigma(s)}(b)\  .
\end{equation}

It is clear that  $\sigma(s)$ is of the following form:
\begin{equation}
\sigma(s)=\left [\begin {array}{ccc} u&0&-v\\\noalign{\medskip}0&1&0
\\\noalign{\medskip}v&0&u\end {array}\right ]\ .
\end{equation}
Then
one readily finds $u={\frac {1}{\sqrt {1+{s}^{2}}}}$
and $v={\frac {s}{\sqrt {1+{s}^{2}}}}$.

This gives
\begin{equation}
b(s)=\left [\begin {array}{ccc} 0&-{\frac {s}{\sqrt {1+{s}^{2}}}}&0
\\\noalign{\medskip}0&0&0\\\noalign{\medskip}0&{\frac {1}{\sqrt {1+{s}
^{2}}}}&0\end {array}\right ]\ .
\end{equation}

Next,  solve
\begin{align}
\frac{\partial b(s,t)}{\partial t}&=[b(s,t)_+,b(s,t)]\\
b(s,0)&=b(s)\ .
\end{align}

 From looking at the power
series
expansion of the solution,   one is led to make 
a guess of the following form:

\begin{equation}
b(s,t)=\left [\begin {array}{ccc} 0&X(t)&Y(t)\\\noalign{\medskip}0&-{\frac
{t }{1+{s}^{2}+{t}^{2}}}&-{\frac {{t}^{2}}{\sqrt {1+{s}^{2}}\left (1+{s}^
{2}+{t}^{2}\right )}}\\\noalign{\medskip}0&{\frac {\sqrt {1+{s}^{2}}}{
1+{s}^{2}+{t}^{2}}}&{\frac {t}{1+{s}^{2}+{t}^{2}}}\end {array}\right ]
\end{equation}
One finds the following solution:
\begin{equation}
b(s,t)=\left [\begin {array}{ccc} 0&-{\frac {s}{\sqrt {1+
{s}^{2}+{t}^{2}}}}&-{\frac {s
t}{\sqrt {1+{s}^{2}}\sqrt {1+{s}^{2}+{t}^{2}}}}\\\noalign{\medskip}0&-{\frac {t}{1+{
s}^{2}+{t}^{2}}}&-{\frac {{t}^{2}}{\sqrt {1+{s}^{2}}\left (1+{s}^{2}+{t}^{2}\right )}}
\\\noalign{\medskip}0&{\frac {\sqrt {1+{s}^{2}}}{1
+{s}^{2}+{t}^{2}}}&{\frac {t}{1+{s}^{2}+{t}^{2}}}\end {array}\right ]
\end{equation}

Finally,  one obtains $a(s,t)$ in the form
\begin{equation}
\label{eq:dress}
a(s,t)=ad_{\tau(s,t)}(a(s))\ ,
\end{equation}
where $\tau(s,t)$ satisfies
\begin{align}
\label{eq;ivp2}
    \frac{d\tau(s,t)}{dt}\tau(s,t)^{-1}&=b(s,t)_+   \\
    \tau(s,0)  &=\sigma(s) \ .
\end{align}

Here is $\tau$:
\begin{equation}\label{dress}
\tau(s,t)=\left [\begin {array}{ccc} {\frac {1}{\sqrt {1+{s}
^{2}}}}&0&-{\frac {s}{\sqrt {1+{s}^{2}}}}\\\noalign{\medskip}-{\frac {s\,t
}{\sqrt {1+{s}^{2}}\sqrt {1+{s
}^{2}+{t}^{2}}}}&{\frac {\sqrt {1+{s}^{2}}}{\sqrt 
{1+{s}^{2}+{t}^{2}}}}&-{\frac {t}{\sqrt {1+{s}^{2}}\sqrt {1+{s}^{2}+{t}^{2}}}}
\\\noalign{\medskip}{\frac {s}{\sqrt {1+{s}^{2}+{t}^{2}}}}&{\frac {t}{\sqrt {1+{s}^{2}+{t}^{2}}}}&{\frac {1}{\sqrt {1+{s}^{2}+{t}
^{2}}}}\end {array}\right ]\end{equation}

Then
\begin{equation}
\kern -50pt a(s,t)=\left [\begin {array}{ccc} -{\frac {s}{1+{s}^{2}}}&{\frac {{s}^{2}t}{\left (1+{
s}^{2}\right )\sqrt {1+{s}^{2}
+{t}^{2}}}}&-{\frac {{s}^{2}}{\sqrt {1+{s}^{2}}\sqrt {1+{s}^{2}+{t}^{2}}}}
\\\noalign{\medskip}-{\frac {t}{\left (1+{s}^{2}
\right )\sqrt {1+{s}^{2}+{t}^{2}}}}&{\frac {{t}^{2
}s}{\left (1+{s}^{2}+{t}^{2}
\right )\left (1+{s}^{2}\right )}}&-{\frac {t
s}{\left (1+{s}^{2}+{t}^{2}
\right )\sqrt {1+{s}^{2}}}}\\\noalign{\medskip}{
\frac {1}{\sqrt {1+{s}^{2}}\sqrt {1+{s}^{2}+{t}^{2}}}}&-{\frac {ts}{\left (1+{
s}^{2}+{t}^{2}\right )\sqrt {1+{s}^{2}}}}&{\frac {s}{1+{s}^{2}+{t}^{2}}}\end {array}\right ]
\end{equation}

Finally, for all $c\in \Frak {sl}_3(\Bbb R)$,  the solution to
\begin{align}
\frac{\partial c(s,t)}{\partial s}&=[a(s,t)_+,c(s,t)]\\
\frac{\partial c(s,t)}{\partial t}&=[b(s,t)_+,c(s,t)]\\
c(0,0)&=c
\end{align}
is given by
\begin{equation}
c(s,t)=ad_{\tau(s,t)}ad_{\sigma(s)}(c)\ .
\end{equation}



{\sc
Department of Mathematics, University of Georgia, Athens, GA
30602}

{\it E-mail address:}
rothstei@@math.uga.edu

\end{document}